\documentclass[journal,12pt,draftclsnofoot,onecolumn,letterpaper]{IEEEtran}
\IEEEoverridecommandlockouts

\usepackage{mathrsfs}
\usepackage{amsfonts}
\usepackage{ifpdf}
\usepackage{cite}
\usepackage{enumerate}

\ifCLASSINFOpdf
\usepackage[pdftex]{graphicx}
\else \fi
\usepackage[cmex10]{amsmath}
\usepackage{array}
\usepackage{mdwmath}
\usepackage{mdwtab}
\usepackage{eqparbox}
\usepackage[tight,footnotesize]{subfigure}
\usepackage{algorithmic}
\usepackage{algorithm}
\usepackage{amsmath}
\usepackage{epsfig}
\usepackage{ae}
\usepackage{amssymb}
\usepackage{times}
\usepackage{amsmath}   
\usepackage{lettrine}
\usepackage{xcolor}

\usepackage{amsthm} 

\usepackage{supertabular} 

\usepackage{multirow}
\newlength\savewidth

\begin{document}

\title{
TransDetector: A Transformer-Based Detector\\for Underwater Acoustic\\Differential OFDM Communications
\thanks{This work was supported in part by the National Science Foundation of China under Grant 61971462 and Grant 61831013; and in part by the State Key Laboratory of Integrated Services Networks (Xidian University) under Grant ISN19-09.}
\thanks{The authors are with the School of Electronic Information and Communications, Huazhong University of Science and Technology, Wuhan 430074, China (e-mail: yuzhouli@hust.edu.cn; wsx@hust.edu.cn; liudime@hust.edu.cn; chuangzhou@hust.edu.cn;)
}
}
\author{
    \IEEEauthorblockN{
        Yuzhou Li,~\IEEEmembership{Member,~IEEE}, Sixiang Wang, Di Liu, and Chuang Zhou\\
    }
    }
\maketitle 
\IEEEpeerreviewmaketitle
\begin{abstract}
  Inter-carrier interference (ICI) and noise mitigation is crucial for precise signal detection in underwater acoustic (UWA) differential orthogonal frequency division multiplexing (DOFDM) communication systems. In this paper, we adopt the Transformer to design a detector, referred to as the TransDetector, which can dramatically mitigate ICI implicitly and noise explicitly, even without requiring any pilot. Compared with the standard Transformer, we come up with three creative designs. Firstly, we break the inner-encoder computation paradigm of the multi-head attention (MHA) in the standard Transformer, and design a brand new inter-encoder attention mechanism, referred to as the interactive MHA, which can significantly improve the performance, as well as accelerate the convergence rate. Secondly, to reduce the noise component attached to the received signal, we design an auto-perception denoising structure, which allows the network to learn the noise distribution in received signals. Thirdly, to better match the characteristics of DOFDM signals and selectively focus on the data at specified locations, we propose a trapezoidal positional encoding (PE), instead of adopting the original sine-cosine PE in the Transformer. Experimental results on both the realistic underwater channel and the simulation channel show that the TransDetector outperforms the classical $\mathcal{X}$-FFT algorithms and the DNNDetector in terms of the BER and the MSE. For example, the BER achieved by the TransDetector is reduced by $27.21\%$ and $12.50\%$ when the signal-to-noise ratio $(\text{SNR})=0$~dB and by $47.44\%$ and $33.49\%$ when $\text{SNR}=20$~dB against the PS-FFT and the DNNDetector based on the realistic channel, respectively.

\end{abstract}
\section{Introduction}
\lettrine[lines=2]{T}{he} underwater acoustic (UWA) differential orthogonal frequency division multiplexing (DOFDM), termed as the UWA-DOFDM, attributed to its attractive features in robustness to multipath effects and near-blind signal detection, has drawn significant attention recently \cite{5606770,8309356,8815817,6840861}. Whereas, achieving precise DOFDM signal detection in complex underwater environments is not easy. Firstly, the low-speed characteristic of the UWA wave (approximately $1500$ m/s) usually results in large Doppler shifts, which would cause severe inter-carrier interference (ICI) to the DOFDM signal\cite{8634811,Li2018}. In addition, the fickle underwater environment noise would further aggravate the distortion of the signal\cite{8634811,Li2018}. Therefore, it is quite important to effectively reduce ICI and noise for precise signal detection in the UWA-DOFDM system.

A variety of ICI mitigation approaches have been proposed in literature, which generally can be divided into two types: the mathematical model (MM)-based and the deep learning (DL)-based. With regard to the MM-based methods, the $\mathcal{X}$-FFT, a category of fast Fourier transform (FFT)-based methods including the P-FFT\cite{5606770,8309356,8815817}, the S-FFT\cite{6840861}, the F-FFT\cite{6489283,ma2016decision}, and the PS-FFT\cite{9500552}, is one of the most typical ones. Specifically, the P-FFT\cite{5606770,8309356,8815817} reduces ICI by slowing down the channel variation through dividing the received time-domain signal into non-overlapping segments, i.e., adding a rectangular window on the received signal. Differently, the S-FFT\cite{6840861} segments the received signal by a raised-cosine window, in which partial overlap between the segmented signals is allowed. The F-FFT\cite{6489283,ma2016decision} suppresses ICI in the frequency domain by compensating Doppler shifts at frequencies of fractions of carrier spacing. Our proposed PS-FFT\cite{9500552} mitigates ICI jointly in the time and frequency domains, i.e., takes advantages of both the P-FFT and the F-FFT.


With regard to the DL-based methods, different kinds of neural networks have been applied to the signal detection problem. Farsad \textit{et al.}\cite{8454325} proposed a recurrent neural network (RNN)-based signal detector and proved that it performs well even without prior-knowledge of the channel state information. A deep neural network (DNN)-based and a convolutional neural network (CNN)-based signal detectors are presented in \cite{2019Deep} and \cite{zhu2021convolutional} for UWA communications, respectively, and it is indicated that both of them outperform the conventional receivers with least-square channel estimation. In addition, it is also feasible to design signal detectors by flexibly combining different types of neural networks. For example, detectors based on the DNN and long short-term memory (LSTM) and on the CNN and bi-directional LSTM were designed in \cite{hassan2022underwater} and \cite{zhang2022deep} for the UWA-OFDM system, respectively.

To summarize, the $\mathcal{X}$-FFT, despite showing significant effectiveness in ICI mitigation, usually suffers three problems. Firstly, the mitigation degree is limited in the case of large Doppler shifts. Secondly, the performance of $\mathcal{X}$-FFT would dramatically deteriorate in low signal-to-noise ratio (SNR) situations, since the noise is not considered in algorithm design. Finally, considerable pilots are required to obtain channel information to enhance the detection precision, which inevitably consumes the quite limited bandwidth, typically only on the order of kHz to tens of kHz\cite{lanbo2008prospects,9427998}. Contrary to the $\mathcal{X}$-FFT, channel estimation is usually not indispensable in DL-based detectors \cite{8454325,2019Deep,zhu2021convolutional}, thus pilot overheads can be substantially reduced. This advantage is extremely attractive for signal detection in UWA communications, where the channel is complex, and even a standardized channel model is not available until now\cite{8610340}.

Based on the analysis above, we in this paper adopt the DL-based method to design a detector for UWA-DOFDM signal detection. Essentially, the process of signal detection is extracting the original transmitted sequences from the received signals polluted by ICI and noise. Interestingly, we find this process is mathematically analogous to the semantic extraction problem in the field of natural language processing (NLP), which aims to extract the target semantic information from a verbose statement. The Transformer\cite{NIPS2017_3f5ee243}, due to its strong global perception ability, has become one of the most competitive networks to extract the semantic information. In addition, it has been shown that the Transformer performs well in various kinds of fields, such as computer vision\cite{https://doi.org/10.48550/arxiv.2010.11929, NEURIPS2021_854d9fca, Zhang_2022_CVPR} and semantic communications\cite{9632815}. Thus, it is interesting to explore the potential of the Transformer in signal detection.

In this paper, we adopt and elegantly modify the standard Transformer to design a detector for the UWA-DOFDM system, referred to as the TransDetector. Especially, to characterize the DOFDM signal and reduce ICI and noise as much as possible, we design a novel positional encoding (PE), a denoising structure, and a new multi-head attention (MHA) scheme. By these efforts, the TransDetector could dramatically mitigate noise explicitly and ICI implicitly, even without requiring any pilot. The experimental results based on both the simulation channel and the realistic underwater channel provided in Watermark\cite{Watermark} show that the mean square error (MSE) and bit error rate (BER) achieved by the TransDetector are significantly smaller than those of the $\mathcal{X}$-FFT and the DNN-based signal detector (DNNDetector).

The main contributions of this work are threefold:

\begin{itemize}
  \item We break the paradigm that the computation of attention in the MHA, designed in the standard Transformer\cite{NIPS2017_3f5ee243}, is limited only within the single layer, and design a brand new attention scheme, referred to as the interactive-MHA (I-MHA), which enables attention matrixes to be flexibly shared across encoder layers. The I-MHA leverages the CNN to fuse the attention matrixes in different layers, by which the network can take full advantage of the semantic information generated at any level, while incurring almost no extra parameter costs. The experimental results show that the I-MHA could tremendously enhance the performance of the network (e.g., reduce the MSE against the baseline network\footnote{The baseline network in this paper refers to the network that adopts the standard Transformer-encoder in \cite{NIPS2017_3f5ee243} with the residual connection replaced by the DeepNorm method proposed in  \cite{https://doi.org/10.48550/arxiv.2203.00555}.} by $12.22\%$ across the test set), as well as significantly accelerates the convergence rate.

  \item Although the implicitly denoising capability might exist in the standard Transformer, we elaborately design an explicitly denoising structure (DNS) to reduce the noise component attached to the received signal. By inserting a feature-mapped additive white Gaussian noise (AWGN) vector into the input matrix, the DNS allows the network to learn the noise component in the received signals by the attention mechanism. The experimental results show that, with the proposed DNS, the MSE of the network against the baseline network is reduced by $10.93\%$ across the test set.

  \item A remarkable feature of the DOFDM signal is that the mutual influence among data sequences carried on subcarriers is weak when they are sufficiently apart. Inspired by this fact, we design a novel PE method, referred to as the trapezoidal-PE (T-PE), instead of adopting the sine-cosine PE method used in the Transformer. The T-PE allows the TransDetector to selectively focus on the data at specified locations, by which attentions on the adjacent subcarriers are intentionally strengthened, and thus improving the detection performance. Experimental results show that, with the proposed T-PE, the MSE of the network against the baseline network is reduced by $4.50\%$ across the test set.

\end{itemize}

The remainder of this paper is organized as follows. In Section~\ref{Section:Problem Transformation}, we introduce the UWA-DOFDM signal detection process and the problem transformation. Section~\ref{Section:TransDetectorStruct} describes the detailed structure of the proposed TransDetector. Section~\ref{Section:SimulationResults} shows the performances comparison between the TransDetector and the $\mathcal{X}$-FFT algorithms in both the simulation channel and the realistic underwater channel provided in Watermark, respectively, and gives the visualized analysis of the attention matrixes. Finally, conclusions are summarized in Section~\ref{Section:Conclusions}.

\section{Problem Description} \label{Section:Problem Transformation}
In this section, we first describe the signal detection process of the UWA-DOFDM system, along with analyzing the factors that affect the accuracy of the detection, and then introduce how to transform the signal detection problem into a problem similar to semantic extraction.

\subsection{Signal Detection for the UWA-DOFDM System}
Considering a UWA-DOFDM symbol containing $K$ sub-carriers with the original data symbol $b_k$ being transmitted on each subcarrier, the final data symbol $d_k$ on the $k$-th subcarrier is obtained by applying the differential encoding\cite{9500552} to the $b_k$, as follows
\begin{equation}\label{Eq:diffCode}
d_{k}=\begin{cases}
\ b_k d_{k-1},& 1\le k\le K-1 \\
\ a_0,& k=0 \\
\end{cases}
\end{equation}
where $a_0$ is a constant known at both the transmitter and the receiver, $b_k$ is generated from the $M$-ary unit-amplitude phase-shift keying (PSK) constellation alphabet set $\mathbb{S}=\left \{ s_{0}, s_{1}, \cdots, s_{M-1} \right \}$, in which the constellation symbol is $s_{m}=e^{j2\pi m/M}, m=0,1, \cdots ,M-1$. The transmitted differential data $d_k$ at the receiver could be modeled by
\begin{equation}\label{Eq:diffDetect}
{{y}_{k}}={{H}_{k}}{{d}_{k}}+\text{ICI}_k+{{n}_{k}}
\end{equation}
where $H_k$, $\text{ICI}_k$, and $n_k$ represent the channel transfer function, ICI, and noise on the $k$-th subcarrier, respectively. Further, by applying signal detection to the received symbol $y_k$ and $y_{k-1}$, the differential estimation of the $k$-th transmitted symbol $\hat{b}_k$ at the receiver is acquired as
\begin{equation}\label{Eq:diffDetect}
{{\hat{b}}_{k}}=\frac{{{y}_{k}}}{{{y}_{k-1}}}=\frac{{{H}_{k}}{{d}_{k}}+\text{ICI}_k+{{n}_{k}}}{{{H}_{k-1}}{{d}_{k-1}}+\text{ICI}_{k-1}+{{n}_{k-1}}}.
\end{equation}
Under the premise that the system bandwidth remains fixed, when the carrier spacing is small enough or the number of sub-carries is large enough, it can be approximated that $H_k\approx H_{k-1}$ holds\cite{5606770}. Because in this case, the channel frequency response between two adjacent carriers can be regarded as changing slowly. In addition, in the case that the noise and ICI are negligible, the $k$-th transmitted symbol $\hat{b}_k$ at the receiver can be approximately obtained by
\begin{equation}\label{Eq:appro_Detect}
{\hat{b}}_{k}\approx \frac{{H}_{k}{d}_{k}}{{H}_{k-1}{d}_{k-1}}\approx\frac{d_k}{d_{k-1}}=b_k.
\end{equation}

In fact, the ICI and noise are usually non-negligible in the underwater environment, even they two are important culprits that cause severe distortion to the received UWA signal, which makes Eq.(\ref{Eq:appro_Detect}) difficult to hold in practice. That is to say, reducing the influence of ICI and noise to a sufficiently small degree is the prerequisite for accurate signal detection of UWA-DOFDM systems.

\subsection{Problem Transformation}
Signal detection aims to extract the original data sequences from the received ones. Both the received data and detection results are time sequences in form, so the procedure of signal detection can be modeled as a kind of sequence-to-sequence model. This procedure is analogous to the semantic extraction problem in the field of NLP, which retrieves the desired semantic information from the source sentence. We can regard the received signal of the UWA-DOFDM system as the source sentence, and the difference detection result is the target semantic information we want, so that these two problems can be considered as equivalent in mathematics. The above facts motivate us to look for inspiration in the field of NLP to deal with the problem of signal detection.

Given the rising popularity of the Transformer\cite{NIPS2017_3f5ee243} as an excellent solution to tackle the semantic extraction problem, we can naturally consider applying it to the signal detection task. In this paper, we devote to designing a Transformer-based detector to reduce both ICI and noise in received signals. However, we cannot simply adopt the Transformer for DOFDM signal detection yet, owing to the significant distinctions between DOFDM signal detection and semantic extraction. Consequently, some elegant modifications are required to match the characteristics of UWA-DOFDM signals, as described in Section~\ref{Section:TransDetectorStruct}.

\section{Design of the TransDetector} \label{Section:TransDetectorStruct}
In this section, we first outline the overall architecture of the TransDetector, then detail the specific modules designed for the TransDetector to match the characteristics of UWA-DOFDM signals.

\subsection{Overall Architecture of the TransDetector}
\begin{figure*}[t]
\centering \epsfxsize=6.5 in \epsfbox{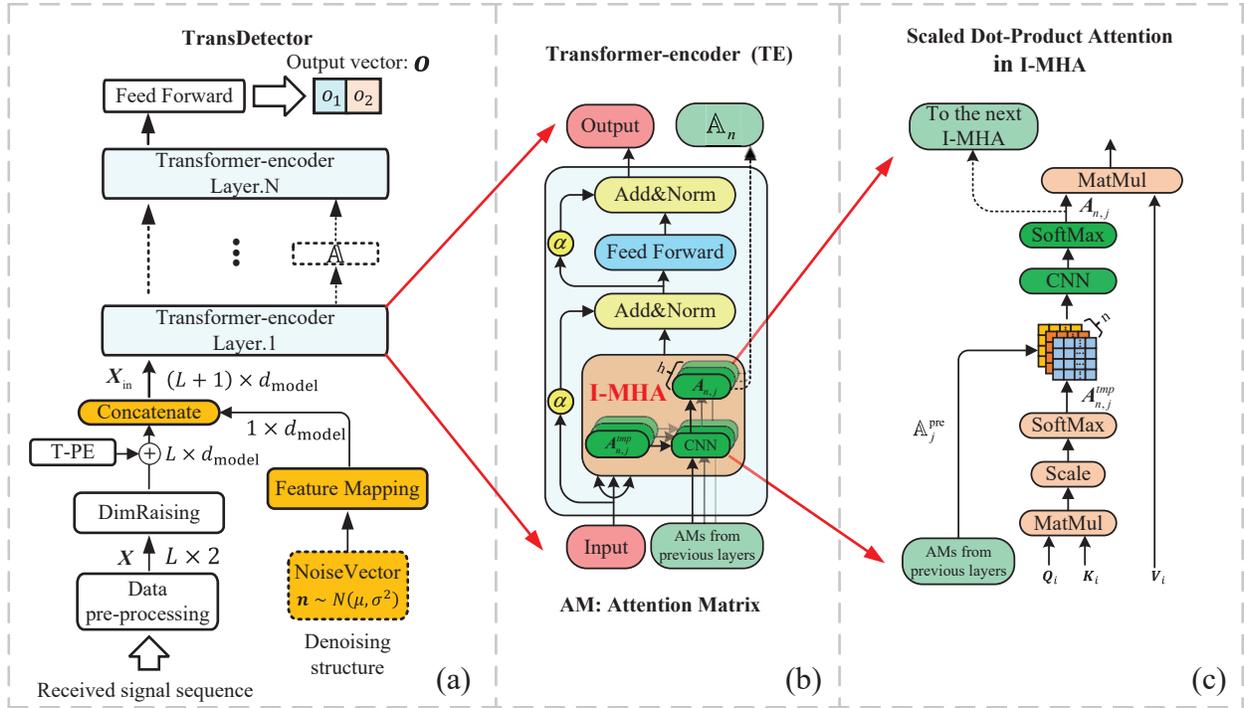}
\centering \caption{The overall architecture of the TransDetector. (a) Data flow diagram of the TransDetector. (b) Illustration of the I-MHA structure in each TE layer. (c) Details of the scaled dot-product attention in the I-MHA.}\label{Fig:TransDetector}
\end{figure*}

The overall architecture of the TransDetector is illustrated in Fig.~\ref{Fig:TransDetector}. The core component of the TransDetector is composed of $N$ Transformer-encoder (TE) layers which have exciting semantic extraction capability for the translation task in NLP. Crucially, the TransDetector also includes several modifications to improve the effectiveness of the design, including a data pre-processing module, the T-PE, the DNS, and the I-MHA. These modules enable the TransDetector to better process the UWA-DOFDM data while allowing it to make more accurate predictions.


At the outset, the input data $\mathbf{X}$ of the TransDetector are generated by a data pre-processing module from received signals. Subsequently, the DNN is utilized by the DimRaising layer to substitute the initial word embedding module for raising the dimension of the input data to $d_\text{model}$. These $d_\text{model}$-dimensional data are further encoded by trapezoidal positional encoding and then pushed into the first TE layer along with a randomly generated $d_{\text{model}}$-dimensional noise vector that follows Gaussian distribution. During the forward propagation of data between different TE layers, instead of using the MHA in the standard Transformer, the I-MHA is used to fuse the attention matrixes of the specified TE layers.

The output layer of the TransDetector is a DNN structure with a tanh activation function. It yields a vector containing two real numbers at one time. These two numbers denote the real and imaginary components of one constellation point in $\mathbb{S}$, respectively. In this paper, quadrature phase shift keying (QPSK) is used to modulate the transmission data, and it is mapped to the constellation points set $\mathbb{S}_{Q}=\left \{ 1+1j , 1-1j, -1-1j, -1+1j \right \}$. Therefore, the output vector of the network is given by
\begin{equation}\label{Eq:result}
\mathbf{o}={f}_\text{TD}\left ( \mathbf{X}  \right ) = \left [ o_1, o_2 \right ] \Rightarrow \hat{b}_k = o_1+jo_2
\end{equation}
where $\mathbf{o}$, ${f}_\text{TD}$, and $\mathbf{X} = \left [ \mathbf{x}_1^T,\mathbf{x}_2^T,\cdots , \mathbf{x}_L^T  \right ] ^T$ represent the output vector, the forward propagation process, and the input matrix of the TransDetector, respectively. In addition, $\mathbf{x}_i,~i=1,2,\cdots,L$, is a two-dimensional vector representing a received complex data in the frequency domain. $o_1$ and $o_2$ represent the real and imaginary parts of the estimated constellation point in $\mathbb{S}_{Q}$, respectively, and they range from $-1$ to $1$.


\subsection{Data Pre-processing}
The following two issues need to be tackled before the TransDetector runs:
\begin{enumerate}[(a)]
\item The initial Transformer is limited to dealing with real data, whereas the data in the UWA communication is usually complex. Thus, it is necessary to preprocess the received complex data so that it can be operated by a network that accepts real inputs.
\item The Transformer cannot output complex data directly and the input of the TransDetector is differential data. Therefore, how to set the output format of the TransDetector corresponding to the input data is also important.
\end{enumerate}

\begin{figure}[t]
\centering \leavevmode \epsfxsize=3.3 in  \epsfbox{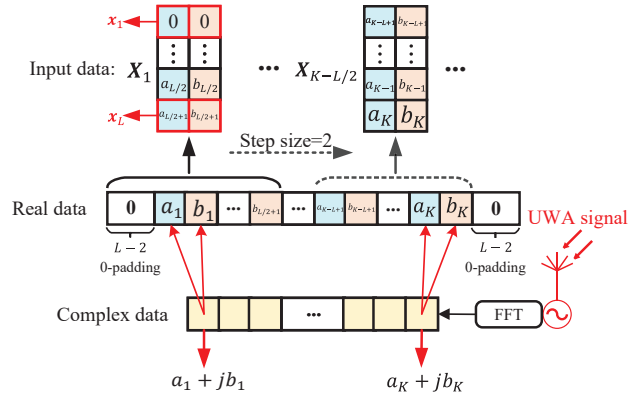}
\centering \caption{Diagram of the pre-processing module for the received data containing $K$ complex numbers.}\label{Fig:data_pre-processing}
\end{figure}

We employ an unfolding method to convert the complex data into a real-valued one, as illustrated in Fig.~\ref{Fig:data_pre-processing}. Specifically, the $K$ received data are first converted to $K$ frequency-domain data by adopting the FFT, and then the real and imaginary components of the obtained frequency domain data are split and sequentially arranged as a sequence containing $2K$ real numbers. Then these $2K$ real numbers are normalized by
\begin{equation}\label{Eq:mnorm}
x_\text{norm}=\frac{x-\bar{x} }{\bar{x}_{10,\text{max}}-\bar{x}_{10,\text{min}} }
\end{equation}
where $\bar{x}$, $\bar{x}_{10,\text{max}}$, and $\bar{x}_{10,\text{min}}$ represent the mean, the mean of the first 10 maximum numbers, and the mean of the first 10 minimum numbers of all $2K$ real numbers, respectively. Besides, the $0$-padding is used to fill the data before and after this real sequence. Subsequently, a sliding window with the window length of $2L$ is utilized to extract the input data from this sequence, i.e., each input of the TransDetector consists of $L$ $2$-dimensional real numbers, which are flattened from $L$ complex numbers (where $L$ is even). This $2L$-dimensional normalized vector is finally reshaped into a matrix with the shape of $L \times 2$ as the input to the TransDetector. In this way, the calculation of complex numbers is rendered equivalent to the calculation of real numbers, ensuring that the network can compute the complex number.

It should be noted that, for convenience of description, the number of input data of the TransDetector will be counted in complex form by default throughout the rest of the paper. We should know that these complex numbers are actually converted to real data before being input into the network.

The TransDetector treats the forward propagation as a regression problem and outputs a vector consisting of two real numbers, representing the real and imaginary parts of the predicted constellation point in $\mathbb{S}$, respectively. In other words, the input-to-output ratio of the TransDetector is $L:1$. Since the input data of the TransDetector are differentially encoded, we set the final output of the network to correspond to the differential detection result of the data $\frac{L}{2}$ and $\frac{L}{2}+1$ in the input, while the remaining input data are used to reduce the interference of ICI on the output. The purpose of establishing this input-to-output relationship will be further discussed in Section~\ref{Section:TransDetectorStruct}-C.

\subsection{Trapezoidal Positional Encoding}
In the Transformer, PE is designed to utilize the order of the input sequence, since the attention mechanism is insensitive to the positional information of the input sequence. PE is also needed in the TransDetector since the differential detection needs to know the position of the front and back data.

Two characteristics need to be considered before designing PE for the TransDetector. Firstly, in the DOFDM system, the ICI leads to mutual interference on the adjacent subcarrier and it is generally observed that, the farther the distance between data sequences transmitted on different subcarriers, the weaker their mutual influence. Second, the TransDetector only produces one detection result for an input containing $L$ complex numbers. This output corresponds to the differential detection result of the two differential data located at the position $\frac{L}{2}$ and $\frac{L}{2}+1$ among the input. In this case, Eq.~(\ref{Eq:result}) can be further explained as
\begin{equation}\label{Eq:result2}
f_\text{TD}\left ( \mathbf{X}  \right ) = \text{est}\left [ \frac{\mathbf{x}_{L/2+1}}{\mathbf{x}_{L/2}} \right ] =\left [ o_1, o_2 \right ] \Rightarrow \hat{b}_k = o_1+jo_2
\end{equation}
where $\text{est}\left [ \frac{\mathbf{x}_{L/2+1}}{\mathbf{x}_{L/2}} \right ]$ represents the differential estimation result for the differential data $\mathbf{x}_{L/2+1}$ and $\mathbf{x}_{L/2}$. The above two factors are important references for our PE design. We hope the network to attend the influence (ICI) level of the surrounding data on the two intermediate data through the attention mechanism and then mitigate the ICI. Especially, the PE function should have equal attention to data on both sides of the input.

\begin{figure}[t]
\centering
  \leavevmode
  \epsfxsize=3 in  \epsfbox{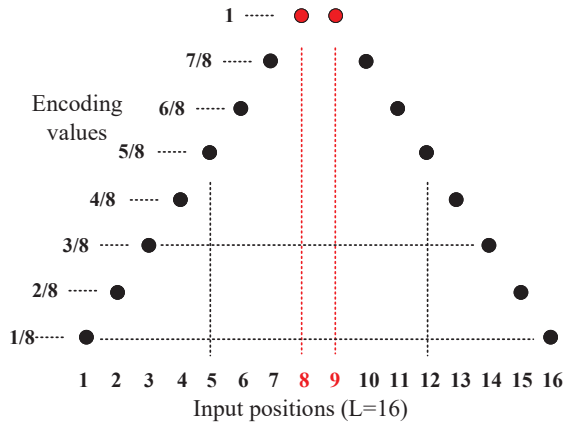}
  \caption{Illustration of the trapezoidal positional encoding.}\label{Fig:position_coding}
\end{figure}

The sine-cosine positional encoding proposed in \cite{NIPS2017_3f5ee243} is inadequate for matching the two characteristics mentioned above. Therefore, we propose the T-PE method to help the network to focus on certain areas of the input data. As shown in Fig.~\ref{Fig:position_coding}, there are $16$ complex data input to the network each time ($L=16$). The encoded values that are added to each input vector after word embedding decrease linearly from the middle to both sides. For example, the constant $1$ is added to the vectors corresponding to the data $L/2=8$ and data $L/2+1=9$, and the constant $7/8$ is added to the vectors corresponding to the data $L/2-1=7$ and data $L/2+2=10$, etc. The relationship between the position of the input vector and the encoding value is given by
\begin{equation}\label{Eq:PE}
\text{PE}_\text{pos}=\text{PE}_{L-\text{pos}+1}=\frac{2}{L} \cdot \text{pos}, ~ 1 \le \text{pos}\le \frac{L}{2}
\end{equation}
where pos represents the absolute position of the current data in all input data.

The T-PE encourages the TransDetector to pay more attention to the data closer to the intermediate two data, since those data points at these locations have a greater influence on the output. The visual analysis about the impact of data present at diverse input locations on the output will be illustrated in Section~\ref{Section:SimulationResults}-C.

\subsection{Denoising Structure}
We pay heed to the fact that the attention mechanism in the Transformer is capable of learning the interaction between input vectors. Drawn by this, we intuitively speculate if it is possible to allow the network to learn the effect of the noise on the received data, thereby attenuating this influence. Inspired by this intuition, we design the DNS to reduce the effect of noise explicitly.

The illustration of the DNS is shown at the bottom right of Fig.~\ref{Fig:TransDetector}(a). Initially, an AWGN vector $\mathbf{n}$ is randomly generated and then mapped to a $d_\text{model}$-dimensional feature vector using a DNN. After that, this feature vector is inserted into the input sequence of the TransDetector, as the input vector for the $(L+1)$-$th$ position. This process is given by
\begin{equation}\label{Eq:denoise}
\mathbf{X}_\text{in}=\text{concat} \left [ \mathbf{X},\text{DNN}\left ( \mathbf{n} \right )  \right ]= \begin{bmatrix}
\mathbf{X} \\
\text{DNN}\left ( \mathbf{n} \right )
\end{bmatrix} _{\left ( {L}+1 \right ) \times d_{\text{model}}}.
\end{equation}
It is expected that the DNS would allow the TransDetector to discover the noise component in the input data and subsequently mitigate it. In this paper, the DNS only acts on the first TE layer of the TransDetector, i.e., the noise vector does not continue to propagate among TE layers after the first layer. The denoising effect of the DNS will be discussed in Section~\ref{Section:SimulationResults}-A.

\subsection{Interactive Multi-Head Attention}
As the information propagates forward through different encoder layers in the TransDetector, the critical semantic information from lower levels is at risk of being lost, which may reduce the overall performance. However, the original Transformer does not take this problem into account. The computation of the MHA in the standard Transformer is limited only within a single encoder layer, we call this inner-encoder. Taking the inner-encoder as a breakthrough point, we design an inter-encoder scheme, referred to as the I-MHA. Different from the MHA, CNN is introduced in the I-MHA for flexibly fusing the attention matrixes across different encoder layers, allowing the shallow semantic information to be used by higher encoder layers. The detailed computation process of the I-MHA is as follows.

As illustrated in Fig.~\ref{Fig:TransDetector}(b) and Fig.~\ref{Fig:TransDetector}(c), the difference between the I-MHA and the MHA mainly lies in the way they calculate the scaled dot-product attention. Specifically, given that the attention matrixes from arbitrary $n$ TE layers will be fused in the TE layer $n$ by the I-MHA and each I-MHA consists of $h$ heads. Firstly, the query, key, and value matrixes $\mathbf{Q}_{i,j}$, $\mathbf{K}_{i,j}$, and $\mathbf{V}_{i,j}$ in the layer $i$ are given by \cite{NIPS2017_3f5ee243}
\begin{align}
{\mathbf{Q}_{i,j}}&=\mathbf{X}_{i}\mathbf{W}_{i,j}^{Q},j=1,2,\cdots,h \notag\\
{\mathbf{K}_{i,j}}&=\mathbf{X}_{i}\mathbf{W}_{i,j}^{K},j=1,2,\cdots,h \\
{\mathbf{V}_{i,j}}&=\mathbf{X}_{i}\mathbf{W}_{i,j}^{V},j=1,2,\cdots,h \notag
\end{align}
where $\mathbf{W}_{i,j}^Q\in \mathbb{R} ^{d_\text{model}\times d_k}$, $\mathbf{W}_{i,j}^K\in \mathbb{R} ^{d_\text{model}\times d_k}$, and $\mathbf{W}_{i,j}^V\in \mathbb{R} ^{d_\text{model}\times d_k}$ identify the linear transformation matrixes of the query, key, and value matrixes of the $j$-$th$ head in the TE layer $i$, respectively, $d_k=d_\text{model}/h$, and $\mathbf{X}_i$ stands for the input matrix of the TE layer $i$. The standard scaled dot-product attention is adopted in the TE layer $i$ to calculate the attention matrix $\mathbf{A}_{i,j}$ of the $j$-th head, which is given by
\begin{equation}\label{Eq:atten_matrix}
\mathbf{A}_{i,j}=\text{softmax}\left ( \frac{\mathbf{Q}_{i,j} \mathbf{K}_{i,j}^T }{\sqrt{d_k} }  \right ),i=1,2,\cdots,n-1
\end{equation}
where $\left ( \cdot  \right ) ^T$ represents the transpose of the matrix. The attention matrixes in all the $j$-th heads of the first $n-1$ TE layers are saved, and finally obtain a set $\mathbb{A}_{j}^\text{pre}$ including $n-1$ attention matrixes as follows
\begin{equation}\label{Eq:Apre}
\mathbb{A}_{j}^\text{pre}=\left \{ \mathbf{A}_{1,j},\mathbf{A}_{2,j},\cdots \mathbf{A}_{\left ( \text{n-1} \right ),j } \right \}, j=1,2,\cdots,h.
\end{equation}

Then, the initial attention matrix $\mathbf{A}_{n,j}^\text{tmp}$ of the head $j$ in the TE layer $n$ is computed by Eq.(\ref{Eq:atten_matrix}). Subsequently, $\mathbf{A}_{i}^\text{tmp}$ is concatenated with the attention matrixes set $\mathbb{A}_{j}^\text{pre}$ to form a new data block, which can be treated as an image data containing $n$ channels. Following this, a CNN layer combined with a softmax is applied to this data block to fuse these $n$ attention matrixes into a new single-channel matrix $\mathbf{A}_{n,j}$, and this process is given by
\begin{equation}\label{Eq:Ai}
\mathbf{A}_{n,j}=\text{softmax}\left \{ \text{CNN}\left [ \text{concat}\left ( \mathbf{A} _{n,j}^\text{tmp}, \mathbb{A}_{n,j}^\text{pre} \right )  \right ]  \right \}.
\end{equation}
This resultant matrix $\mathbf{A}_{n,j}$ is considered as the final attention matrix of the head $j$ in the TE layer $n$. The attention matrixes of the remaining $h-1$ heads in the TE layer $n$ are obtained following the same process. And the attention matrixes $\mathbf{A}_{n,j},j=1,2,\cdots,h$ in the current $n$-th TE layer can be further utilized by the next I-MHA in the subsequent TE layers for fusion.

\subsection{Layer Normalization and Activation Function}
The layer normalization (LN) used in the original Transformer\cite{NIPS2017_3f5ee243} is termed as the Post-LN. Ruibin \textit{et al.}\cite{https://doi.org/10.48550/arxiv.2002.04745} found that this standard LN method would lead to unstable gradient distribution between different layers, and thus proposed a Pre-LN method to improve the stability of the Transformer. Unfortunately, the Pre-LN method may reduce the performance of the Transformer. Wang \textit{et al}. \cite{https://doi.org/10.48550/arxiv.2203.00555} proposed a DeepNorm that combines the well performance of Post-LN and the stable training of Pre-LN. It scales the residual connection before normalization, i.e., $x_{l+1}=\text{LN}(\alpha x+f(x))$, where $\alpha$ represents the scaling factor, and when $\alpha=1$, DeepNorm is equivalent to Post-LN. This paper adopts the DeepNorm method for normalization. We use the Xavier initialization method to initialize the parameters of each layer of the TransDetector before training, and the specific gain parameters are referred to \cite{https://doi.org/10.48550/arxiv.2203.00555}.

Different from text information that are usually positive value, UWA-DOFDM signals usually contain a large number of negative values. Therefore, to ensure the negative data can be propagated forward in the model, the Mish activation function\cite{misra2019mish} is used in this paper, as shown below
\begin{equation}\label{Eq:Mish}
\text{Mish}(x)=x\cdot \text{tanh}\left [ \text{ln}\left ( 1+e^{x}  \right )  \right ]
\end{equation}
where $\text{tanh}(x)=(e^x-e^{-x})/(e^x+e^{-x})$. The activation function allows negative numbers to have a slight gradient propagation, rather than a zero boundary as in ReLU \cite{glorot2011deep}. The smooth activation function allows information to penetrate into the neural network better, resulting in better accuracy and generalization.

\section{Experimental Results and Analysis} \label{Section:SimulationResults}
In this section, we evaluate the performance of the TransDetector against the DNNDetector and $\mathcal{X}$-FFT algorithms based on two kinds of UWA channels, including the realistic and the simulation channel, in terms of MSE and BER. Additionally, we explain why the TransDetector can implicitly remove ICI by visualizing the attention matrixes of the TransDetector.


\subsection{Evaluation Based on the Simulation Channel}
In the following simulation experiments, we consider an UWA channel with typical parameter settings provided in\cite{9500552}. The detailed channel and UWA-DOFDM system parameter settings are summarized in Table~\ref{Tab: setting} and Table~\ref{Tab: ofdmsetting}, respectively. 250 consecutive pilot symbols will be inserted into the beginning of each DOFDM data frame as the training data of the combined weight update stage for the $\mathcal{X}$-FFT algorithms involved in this section.

\begin{table}
\centering
\caption{PARAMETER SETTINGS IN THE SIMULATION CHANNEL} \label{Tab: setting}
\begin{tabular}{c|c}
  \hline Water depth & 15~m \\
  \hline Transmitter height & 7.5~m \\
  \hline Receiver height & 7.5~m \\
  \hline Horizontal distance & 2~km \\
  \hline Sound speed in water & 1500~m/s \\
  \hline Sound speed in bottom & 1400~m/s \\
  \hline Spread factor & 1.5 \\
  \hline Number of paths & 19 \\
  \hline
\end{tabular}
\end{table}

\begin{table}
\centering
\caption{PARAMETER SETTINGS FOR UWA-DOFDM SYSTEMS} \label{Tab: ofdmsetting}
\begin{tabular}{c|c}
  \hline Number of subcarriers & 1024 \\
  \hline Number of symbols~/~frame & 8 \\
  \hline Channel bandwidth & 12~kHz \\
  \hline Sampling rate & 192~kHz \\
  \hline Modulation type & QPSK \\
  \hline Guard interval & ZP~(16~ms) \\
  \hline Center frequency & 32~kHz \\
  \hline Doppler factor & $(1\sim 5) \times 10^{-4}$ \\
  \hline SNR & $(0\sim 30)$~dB \\
  \hline
\end{tabular}
\end{table}

We train the TransDetector on the dataset consisting of a total of 50,000 DOFDM frames, which are generated under the condition of $\text{SNR}=\left[ 0,5,10,15,20,25,30 \right]$ and Doppler factor $\alpha=\left[ 1,1.5,2,2.5,3,4,5 \right]\times 10^{-4}$. $80\%$ of these DOFDM frames are randomly selected as the training set, $10\%$ as the verification set, and $10\%$ as the test set.

For the parameter settings of the TransDetector, the output dimension of the DimRaising layer is set to $d_\text{model}=64$, the length of data input to the model is set to $L=16$, the number of heads in each I-MHA is $4$, and a total of $6$ TE layers are used. The I-MHA is implemented in two groups of TE layers, including the TE layers \{$1$, $3$, $4$\} and \{$2$, $5$, $6$\}. The size of the convolution kernel in the I-MHA is set to $3\times 3$, the step size is set to $1$, and the bias is not required. Adam is chosen as the optimizer for the training stage. The update scheme of the learning rate is consistent with \cite{NIPS2017_3f5ee243}, which is given by
\begin{equation}\label{Eq:lr}
\text{lrate} =d_{\text{model}}^{-0.5} \cdot \text{min} \left ( \text{steps}^{-0.5}, \text{steps} \cdot \text{warmup}^{-1.5} \right )
\end{equation}
where $\text{warmup}=\text{2000}$. Mean square error is used as the loss function in the network training stage, which is given by
\begin{equation}\label{Eq:mseloss}
\text{Loss}\left ( \hat{b}_k \right ) =\frac{1}{2n} \sum_{k=1}^{n} \left | \hat{b}_k -b_k\right | ^2.
\end{equation}
Finally, we trained the TransDetector on a server with 4 NVIDIA RTX 2080 Ti GPUs for 10 epochs.

The design of the DNNDetector refers to the literature \cite{2019Deep}, and some specific modifications are made. Total $6$ cascaded fully-connected (FC) layers are adopted by the DNNDetector and all bias parameters in these FC layers are enabled. The detailed settings of these FC layers are shown in Table~\ref{Tab: DNNDetector}. Both the data preprocessing method and output format of the DNNDetector are consistent with the TransDetector. In addition, this section ensures that the numbers of trainable parameters of the DNNDetector and the TransDetector are approximately identical. We use the summary function in the PyTorch-torchsummary library to count the number of trainable parameters of these two networks. When the input data length $L=16$, the total number of trainable parameters of the DNNDetector and TransDetector are 331394 and 313290, respectively.

\begin{table}
\centering
\caption{SETTINGS OF FC LAYERS IN THE DNNDETECTOR} \label{Tab: DNNDetector}
\begin{tabular}{c|c|c|c}
  \hline Layers & Input neures & Output neures & Activation functions  \\
  \hline 1st~(input layer) & $2L$ & 128 & Mish  \\
  \hline 2nd & 128 & 256 & Mish  \\
  \hline 3rd & 256 & 512 & Mish  \\
  \hline 4th & 512 & 256 & Mish  \\
  \hline 5th & 256 & 128 & Mish \\
  \hline 6th~(output layer) & 128 & 2 & Tanh  \\
  \hline
\end{tabular}
\end{table}

\subsubsection{Bit Error Rate}
\begin{figure}[t]
\centering
  \leavevmode
  \epsfxsize=3 in \epsfbox{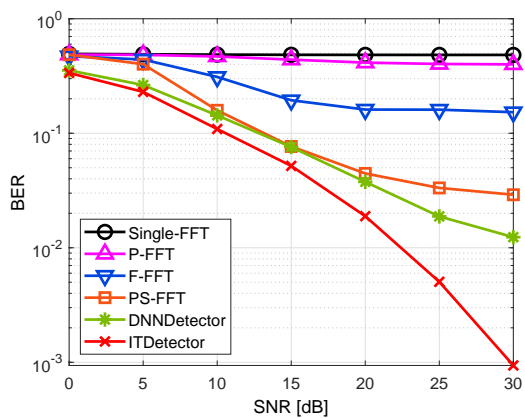}
  \caption{Comparison of the TransDetector, $\mathcal{X}$-FFT, and the DNNDetector algorithms in terms of the BER with different SNR, where the Doppler factor $\alpha~=~3~\times ~10^{-4}$. } \label{Fig:doppler}
\end{figure}

Fig.~\ref{Fig:doppler} compares the BER curves achieved by the TransDetector, $\mathcal{X}$-FFT, and the DNNDetector, respectively, and the Doppler factor $\alpha=3\times10^{-4}$. It can be observed that the Single-FFT without ICI mitigation and the P-FFT algorithms nearly fail to work at this higher Doppler factor. The BER performance of the TransDetector outperforms all the $\mathcal{X}$-FFT and DNNDetector algorithms for the given SNR. Compared with the DNNDetector and the best algorithm PS-FFT in $\mathcal{X}$-FFT, the BER of the TransDetector is reduced by $5.01\%$, $31.22\%$ when $\text{SNR}=0$ dB, and by $42.01\%$, $48.21\%$ when $\text{SNR}=20$ dB, respectively. Moreover, it can be seen that with the increase of SNR, the superiority of the TransDetector becomes more obvious.

\begin{figure}[t]
\centering
  \leavevmode
  \epsfxsize=3 in  \epsfbox{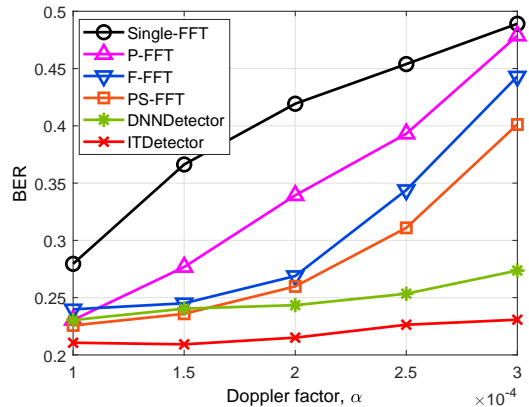}
  \caption{Comparison of different algorithms in terms of the BER with different Doppler factors, where $\text{SNR}=5$~dB.} \label{Fig:snr}
\end{figure}

Fig.~\ref{Fig:snr} shows the BER curves of different algorithms under the $\text{SNR}=5$~dB and the Doppler factor $\alpha$ ranges from $1\times10^{-4}$ to $3\times10^{-4}$. The results show that, the BER of all these algorithms tend to deteriorate as the Doppler factor increases. The BER performance of the TransDetector is still the best among them. When the Doppler factor is less than $1.5\times10^{-4}$, the BER achieved by the DNNDetector is slightly worse than that of the PS-FFT. Compared with the PS-FFT and the DNNDetector, the BER of the TransDetector is reduced by $8.01\%$, $9.88\%$ when Doppler factor $\alpha=1\times10^{-4}$, and by $44.22\%$, $18.23\%$ when $\alpha=3\times10^{-4}$, respectively.

\begin{figure}[t]
\centering
  \leavevmode
  \epsfxsize=3 in  \epsfbox{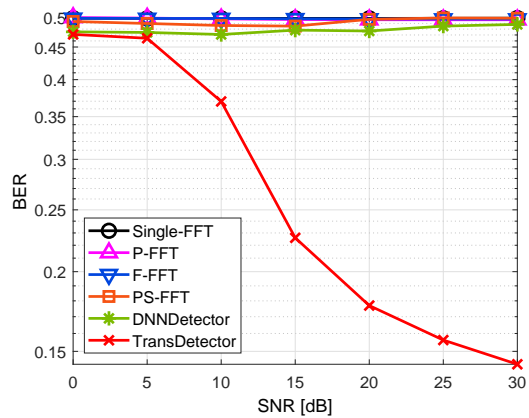}
  \caption{Comparison of different algorithms in terms of the BER with different SNR, where the Doppler factor $\alpha~=~5~\times ~10^{-4}$.} \label{Fig:highdop}
\end{figure}

Fig.~\ref{Fig:highdop} shows the BER curves in the more serious Doppler factor condition, where $\alpha~=~5~\times ~10^{-4}$. Both $\mathcal{X}$-FFT and the DNNDetector algorithms cross the critical point of performance breakdown. The BER obtained by them remains at around $0.5$ all the time, indicating they fail to work. What is striking is that the TransDetector can still achieve fine signal detection performance under such circumstances. When $\text{SNR}=30$~dB, the BER achieved by the TransDetector is $0.146$. TransDetector shows certain ICI mitigation abilities even at high Doppler factors.

\subsubsection{Mean Square Error}
The MSE measures the error between the detected data and transmitted data in terms of dB, described as follows
\begin{equation}\label{Eq:MSE}
\text{MSE}(\hat{b})= 10\log_{10}{ \frac{ \sum_{i=1}^{n} \left | \hat{b}_i-b_i  \right |^2 }{n} }
\end{equation}
where $\hat{b}_i$ is the detection result of the transmitted signal $b_i$.

\begin{figure}[t]
\centering
  \leavevmode
  \epsfxsize=3 in  \epsfbox{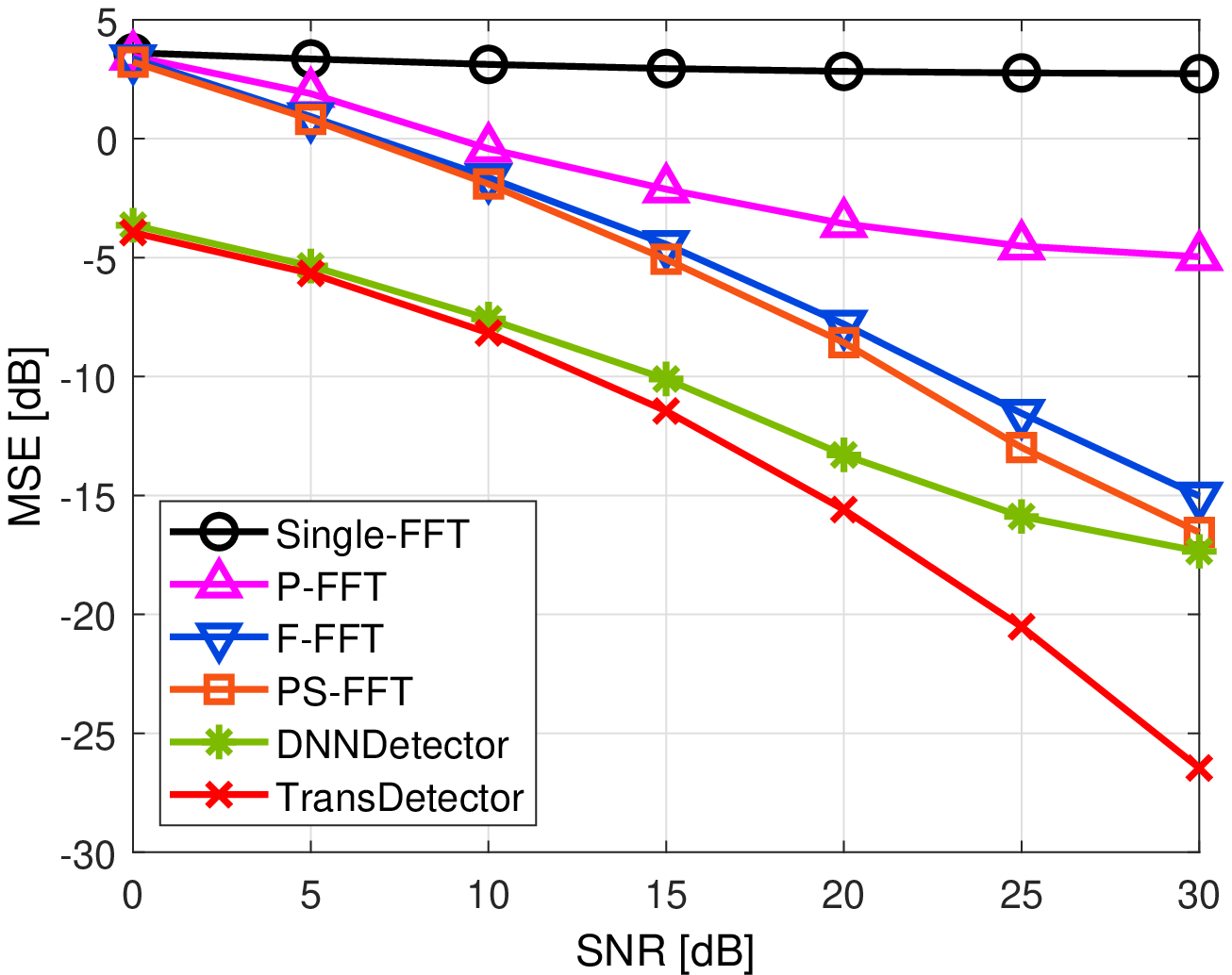}
  \caption{Comparison of different algorithms in terms of the MSE with different SNR, where the Doppler factor $\alpha=2\times10^{-4}$.}\label{Fig:dop_2}
\end{figure}

Fig.~\ref{Fig:dop_2} shows the MSE curves of different algorithms changing with different SNR when the Doppler factor $\alpha=2\times10^{-4}$. It can be seen that the MSE of the TransDetector is significantly better than those of $\mathcal{X}$-FFT algorithms and the DNNDetector, and this advantage becomes more obvious as the SNR increases. Compared with the DNNdetector, the MSE of TransDetector is reduced by $6.88\%$ and $41.10\%$ when $\text{SNR}=0$~dB and $\text{SNR}=20$~dB, respectively.

\subsubsection{Ablation Study of the DNS, the T-PE, and the I-MHA}
Here, we use the MSE to evaluate the effect of the DNS, the T-PE, and the I-MHA on the baseline network.

\begin{figure}[t]
\centering
  \leavevmode
  \epsfxsize=3.1 in  \epsfbox{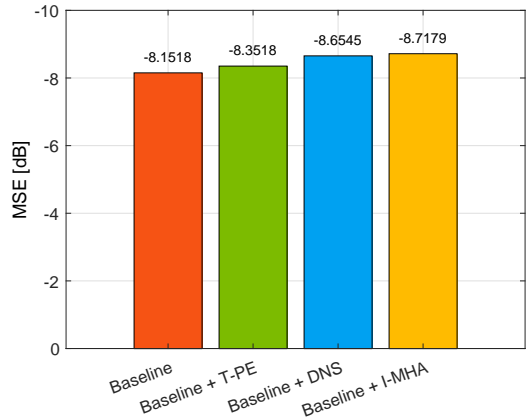}
  \caption{The performance effect on the TransDetector-baseline when using the T-PE, the DNS, and the I-MHA, respectively.}\label{Fig:compare_struct}
\end{figure}

Fig.~\ref{Fig:compare_struct} shows the performance improvement effect when the DNS, the T-PE, and the I-MHA modules are incorporated with the Transdetector-baseline, respectively. The results are calculated on the overall test set and are not limited to a certain SNR or Doppler factor. The baseline version of the TransDetector uses only the original Transformer-encoder structure, without these three proposed modules. It is notable that all these three modules can improve the MSE performance of the baseline network, among which the I-MHA is the most effective. And compared with the standard MHA, the I-MHA used in the experiment consumes only $216$ more parameters, which is almost negligible compared to the overall parameter amount of about $300K$. When the T-PE, the DNS, and the I-MHA are used separately in the baseline network, the overall MSE performance of the baseline network is improved by $4.50\%$, $10.93\%$, and $12.22\%$, respectively, which fully verifies the validity of the three structures designed in this paper. The effect of the T-PE will be further explained in Section.~\ref{Section:SimulationResults}-C.

During the experiment, we find that the distribution of the random noise vector generated in the DNS directly affects the denoising performance of the DNS, and the closer the distribution of the noise vector to the real noise distribution in the received signal, the better the denoising effect.

\begin{figure}[t]
\centering
  \leavevmode
  \epsfxsize=3 in  \epsfbox{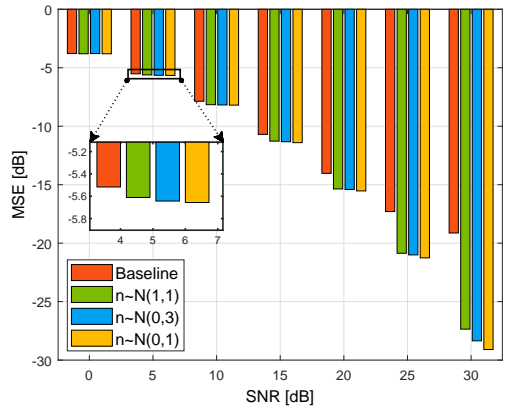}
  \caption{Different optimization effects on baseline when noise vectors used in the DNS obey three different kinds of distributions.} \label{Fig:dns}
\end{figure}

Fig.~\ref{Fig:dns} illustrates the influence of noise vectors with three different distributions on the denoising performance of the DNS when the $\text{SNR}$ ranges from 0 to 30 dB, and these results are obtained on the overall test set. Here, three kinds of DNSs using different noise distributions, i.e., $\mathbf{n} \sim \mathcal{N} \left ( 0,1 \right ) $, $\mathbf{n} \sim \mathcal{N} \left ( 1,1 \right ) $, and $\mathbf{n} \sim \mathcal{N} \left ( 0,3 \right ) $, are applied to the baseline network respectively, and we call these three networks the baseline-DNS1, the baseline-DNS2, and the baseline-DNS3 respectively. Obviously, all these three networks can significantly reduce the MSE of the baseline network. The baseline-DNS1 adopts a noise vector obeying the $\mathbf{n} \sim \mathcal{N} \left ( 0,1 \right ) $ distribution, showing the best MSE performance. When $\text{SNR}=15$~dB, the MSE of the baseline-DNS1 is $3.03\%$, $1.84\%$ and $15.04\%$ lower than that of the baseline-DNS2, the baseline-DNS3, and the baseline, respectively, and the performance gaps reach the maximum when $\text{SNR}=30$~dB. The reason for the above phenomenon is that, in the simulation stage, the noise attached to the received signal is mainly based on an AWGN obeying $\mathbf{n} \sim \mathcal{N} \left ( 0,1 \right ) $. And the original intention of the DNS design is to introduce a specific noise vector, making the network learn the characteristics of this kind of noise vector and obtain the noise component on the received signal through the attention mechanism, and then reduce the noise. As a result, although the other two AWGN vectors obeying other distributions are also effective, the denoising effect is limited.

Therefore, it is important for the DNS that the generated random noise obey as much as possible the same distribution as the noise in the received signal. And how to obtain the noise distribution in the real received signal is also an open problem for follow-up research.


\begin{figure}[t]
\centering
  \leavevmode
  \epsfxsize=3 in  \epsfbox{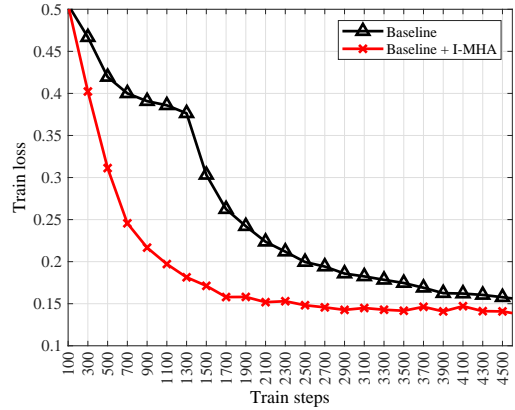}
  \caption{Train loss comparison of the baselines with and without the I-MHA, only the first 4600 steps of the training process are shown here.} \label{Fig:shoulian}
\end{figure}

Fig.~\ref{Fig:shoulian} shows the comparison of the training loss for the first $4600$ steps of baseline with and without the I-MHA. It is evident that the model equipped with the I-MHA converges faster and is close to convergence at step $2000$, while the model without the I-MHA is close to convergence at step $4500$. The loss value of the network using the I-MHA is consistently lower than that of the baseline throughout the training cycle. It is apparent that the I-MHA can significantly accelerate the training speed of the TransDetector.

\subsubsection{Effect of the Receptive Field Size}
\begin{figure}[t]
\centering
  \leavevmode
  \epsfxsize=3.1 in  \epsfbox{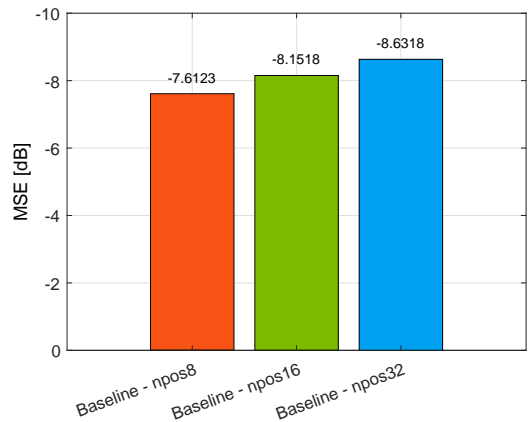}
  \caption{Comparison of the TransDetector under different receptive fields in terms of the MSE, where baseline-npos8 represents a baseline network with a receptive field size of 8.} \label{Fig:inputlength}
\end{figure}

The amount of data input to the network controls the size of the receptive field of the network. We use the sliding window size $L$ mentioned in Section~\ref{Section:TransDetectorStruct}-B to control the receptive field in the TransDetector-baseline. Fig.~\ref{Fig:inputlength} compares the MSE when different receptive fields are adopted by the baseline, in which the receptive field sizes $L=\left \{8,16,32 \right \}$. The results are obtained on the overall test set. Obviously, with the increase of the baseline network receptive field, the overall MSE performance also improves. This trend is also in line with our intuition, that is, the increase in the amount of input data allows the network to pay attention to more interference sources that affect the output, which is more conducive to the network reducing the interference of ICI. The MSE of the baseline-npos16 and baseline-npos32 is $11.68\%$ and $20.92\%$ lower than that of the baseline-npos8, respectively.



\subsection{Evaluation Based on the Realistic UWA Channel}
Watermark provides an effective test platform for the physical layer scheme of underwater acoustic communication. It facilitates researchers to verify, examine, and analyze algorithms for the physical layer under highly realistic and reproducible circumstances\cite{Watermark}.

In this work, we use the NOF1 channel in the Watermark for performance evaluation. NOF1 is a SISO channel measured in the shallow part of Oslofjorden, Norway, and the transmitter and receiver in this scenario are both at the bottom of the water. The data in NOF1 consists of a 32.9-s channel estimate, repeated 60 times at 400-s intervals, yielding a total simulation time of about 33 minutes\cite{Watermark}. The fitted channel information is saved in the form of a matrix, and we can use these data to generate a dataset close to the real environment.

Before applying NOF1, some DOFDM parameters need to be modified to match the characteristics of the NOF1 channel, which are shown in Table~\ref{Tab: NOF1setting}. We use the model trained under the aforementioned simulation channel as the pre-trained network and then train another epoch using the data set generated under the NOF1 channel for fine-tuning.

\begin{figure}[t]
\centering
  \leavevmode
  \epsfxsize=3 in  \epsfbox{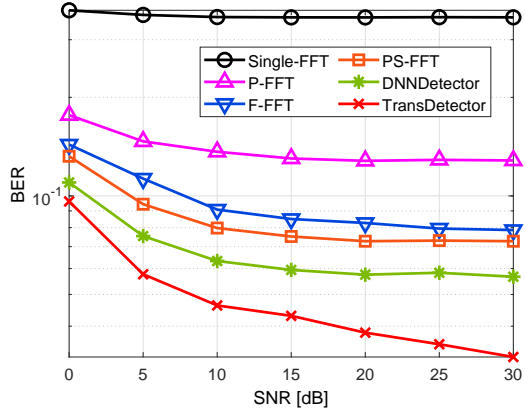}
  \caption{Comparison of different algorithms under the NOF1 channel in terms of the BER.} \label{Fig:NOF1BER}
\end{figure}

Fig.~\ref{Fig:NOF1BER} shows the BER curves comparison of the TransDetector, the DNNDetector, and the $\mathcal{X}$-FFT algorithms under the NOF1 channel. It can be seen that the BER variation trend of the TransDetector algorithm in the NOF1 channel is similar to that in the simulation channel. With the increase of SNR, the BER of all algorithms tend to decrease. Under the channel conditions of $\text{SNR}=0$ dB and $\text{SNR}=20$ dB, the BER achieved by the TransDetector algorithm is lower $27.21\%$ and $47.44\%$ than that of the best $\mathcal{X}$-FFT algorithm, PS-FFT, and lower $12.50\%$ and $33.49\%$ than that of the DNNDetector, respectively. These experimental results show that the TransDetector algorithm still shows strong competitiveness in the real marine environment.

\begin{table}
\centering
\caption{DOFDM PARAMETER SETTINGS IN THE NOF1 CHANNEL} \label{Tab: NOF1setting}
\begin{tabular}{c|c}
  \hline Number of subcarriers & 1024 \\
  \hline Number of symbols~/~frame & 8 \\
  \hline Channel bandwidth & 8~kHz \\
  \hline Center frequency & 14~kHz \\
  \hline Sampling rate & 96~kHz \\
  \hline Guard interval & ZP~(12~ms) \\
  \hline SNR & $(0\sim 30)$~dB \\
  \hline
\end{tabular}
\end{table}

%
%

\begin{figure}[t]
\centering
  \leavevmode
  \epsfxsize=6.5 in  \epsfbox{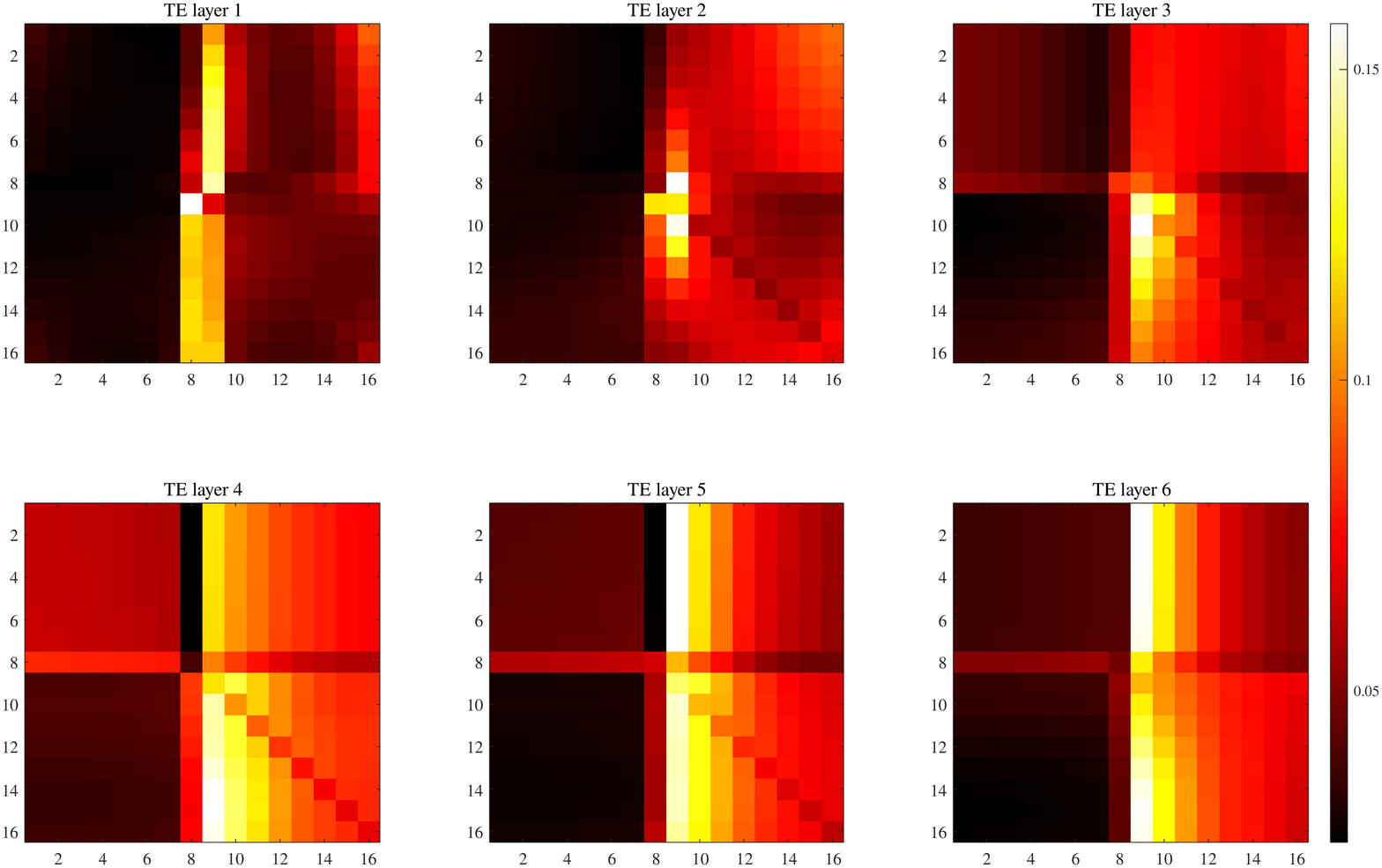}
  \caption{Distribution of the attention matrixes when the TransDetector detects the $1st$ received data in DOFDM symbols. The first $7$ pieces of input data are filled with $0$-padding.} \label{Fig:attmatrix1}
\end{figure}

\begin{figure}[t]
\centering
  \leavevmode
  \epsfxsize=6.5 in  \epsfbox{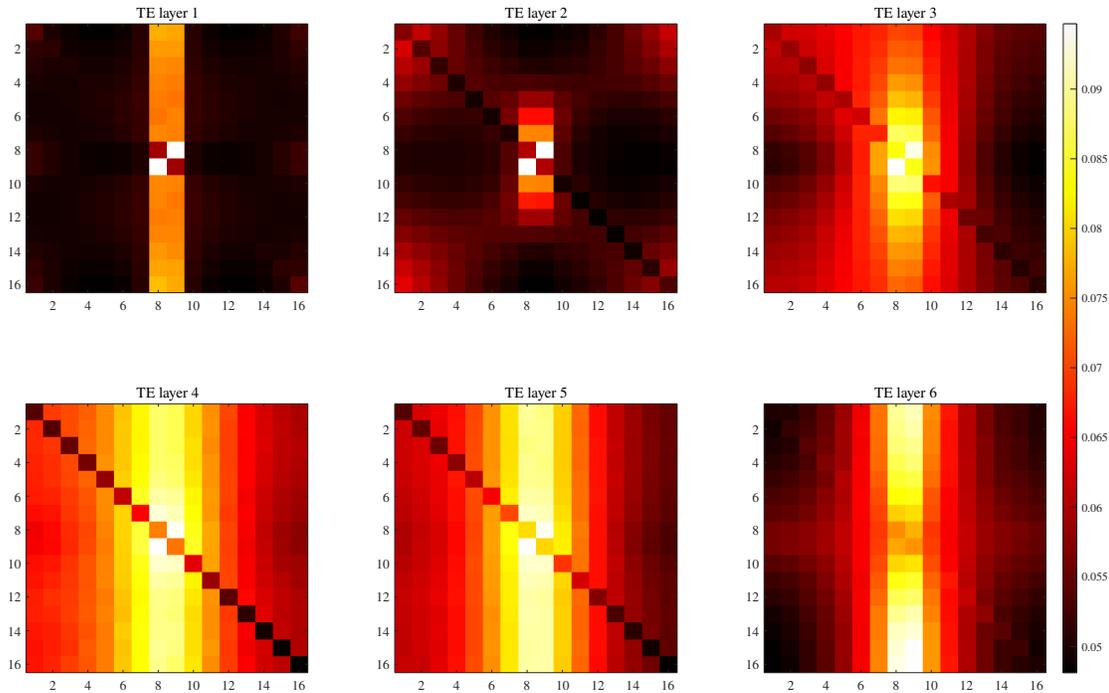}
  \caption{Distribution of the attention matrixes when the TransDetector detects the $512st$ received data in DOFDM symbols.} \label{Fig:attmatrix512}
\end{figure}

\begin{figure}[t]
\centering
  \leavevmode
  \epsfxsize=6.5 in  \epsfbox{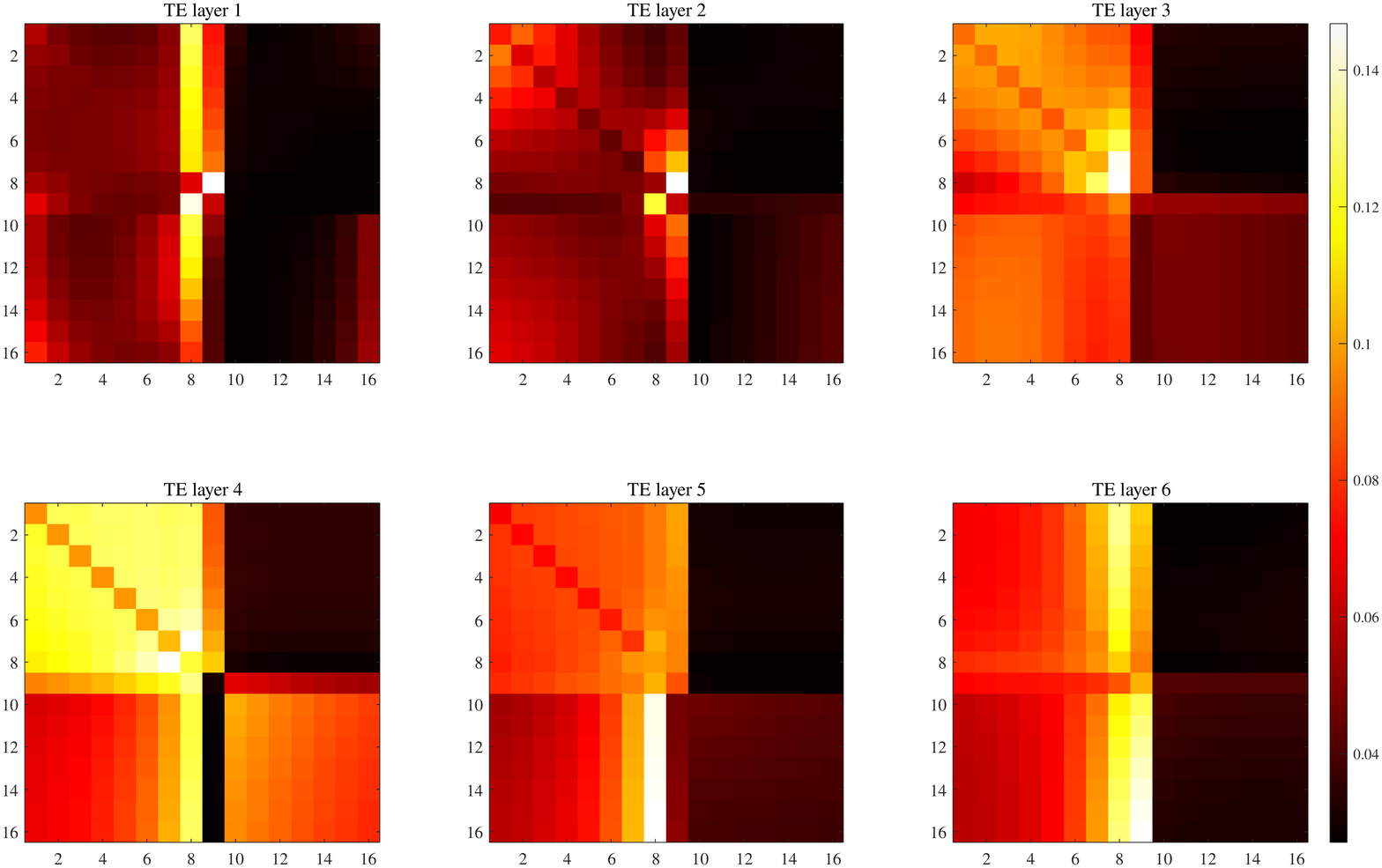}
  \caption{Distribution of the attention matrixes when the TransDetector detects the $1023rd$ received data in DOFDM symbols. The last $7$ pieces of input data are filled with $0$-padding.} \label{Fig:attmatrix1023}
\end{figure}

\begin{figure}[t]
\centering
  \leavevmode
  \epsfxsize=6.5 in  \epsfbox{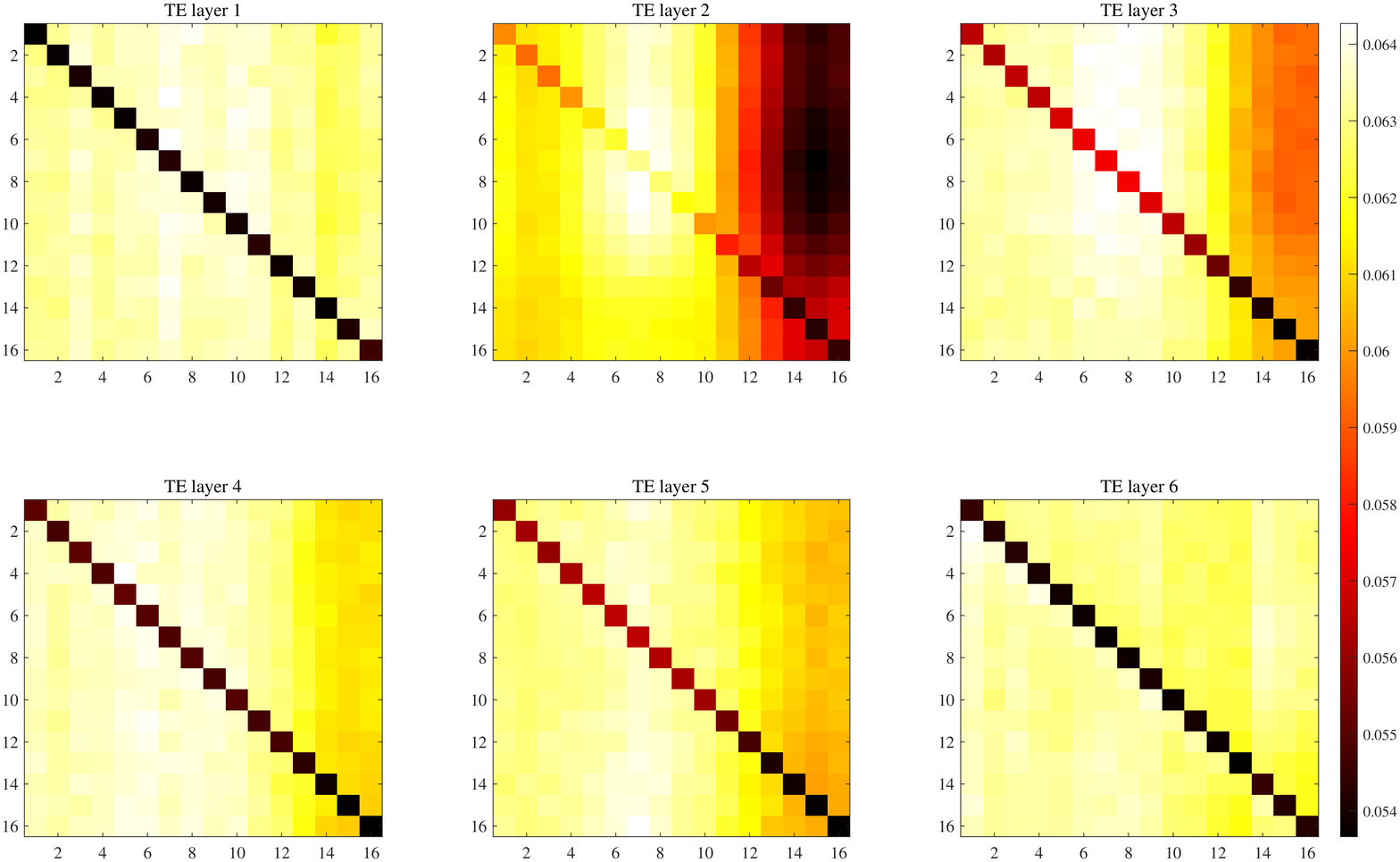}
  \caption{Distribution of the attention matrixes when the TransDetector with no T-PE detects the $512st$ received data in DOFDM symbols.} \label{Fig:attmatrix512_noPE}
\end{figure}

\subsection{Visualization of the Attention Mechanism}
As previously mentioned, ICI is particularly serious in UWA channels due to the low-speed characteristic of the UWA wave, and it has become one of the most important factors affecting the detection accuracy of UWA-DOFDM signals. The existing $\mathcal{X}$-FFT algorithms \cite{5606770,8309356,8815817,6840861,6489283,ma2016decision,9500552} all take the ICI mitigation as the primary task, ignoring the influence of noise, which ultimately leads to limited performance. By utilizing the TransDetector, a detection-focused DL model that leverages the attention mechanism, higher accuracy of signal detection can be accomplished. In this section, we visualize the attention matrixes of the TransDetector to further illustrate why the TransDetector can implicitly reduce ICI.

Fig.~\ref{Fig:attmatrix1} $\sim$ Fig.~\ref{Fig:attmatrix512_noPE} show the distribution of attention values at each TE layer of the TransDetector during the detection of the first ($1st$), the middle ($512th$), and the last ($1023rd$) differential data in a received DOFDM symbol. Each matrix shown in Fig.~\ref{Fig:attmatrix1} $\sim$ Fig.~\ref{Fig:attmatrix512_noPE} is given by
\begin{equation}\label{Eq:attn}
\mathbf{A}_{\text{layer}_i}^{\text{idx}} =\frac{\sum_{n=1}^{N} \sum_{j=1}^{n_{\text{heads}}} \left [ \mathbf{A}_{\text{layer}_i}^{\text{head}_j} \right ]_n^{\text{idx}}}{N}
\end{equation}
where $\mathbf{A}_{\text{layer}_i}^{\text{idx}}$ represents the average values of the multi-head attention matrixes of the TE layer $i$ when detecting the $idx$-$th$ received data in a DOFDM symbol. $n_\text{heads}$ represents the number of the heads in the I-MHA. $\mathbf{A}_{\text{layer}_i}^{\text{head}_j}$ represents the attention matrixe of the head $j$ in the TE layer $i$. And the notation $\left [ \cdot \right ]_n^\text{idx}$ represents the detection of the $\text{idx}$-$th$ data in the $n$-$th$ DOFDM symbol, and we set $N=5000$ in this experiment. Lighter pixels in the matrix reflect the larger values, which indicates that the TransDetector pays more attention to those sites. The elements of $\alpha _{i,j}, ~ 0\le i,j \le L$, in each attention matrix represent the correlation degree between the data at the position $j$ and the data at the position $i$.

Fig.~\ref{Fig:attmatrix512} shows the averaged attention matrixes after detecting the $512th$ data of all $N$ DOFDM symbols. It is evident that the TransDetector pays the most attention to the relationship between the two data in the middle (at the $8th$ and the $9th$ positions) of the input data and the least attention to the data themself (as seen from the diagonals) since each piece of data has no effect on itself. This is in accordance with our settings that the output of the TransDetector is related to the differential detection results of the two middle input data, and is quite consistent with the process of DOFDM signal detection. In Fig.~\ref{Fig:attmatrix1}, the first seven data are filled with 0-padding since there is no data before the first data of the DOFDM symbol. It can be clearly found that the front of all six matrixes is almost black (with very small values), which corresponds to the fact that the $0$-padding is merely used to fill in the input data and has no effect on the output. It turns out that the network really ignores these $0$-padding data. The same conclusion can be seen from Fig.~\ref{Fig:attmatrix1023}.

Fig.~\ref{Fig:attmatrix512_noPE} shows the attention matrixes of the TransDetector with no T-PE. The difference from the first three figures is obvious that, the TransDetector does not seem to be able to focus well on location-specific data without T-PE. Therefore, the T-PE can prompt the network to grasp the positional relationship of the input data through the preset encoding value, thus improving the performance of the network.

Additionally, it can be seen from the first three figures that the values of the pixels farther away from the middle position tend to decrease gradually. It also agrees with the general phenomenon of ICI that the farther the distance between data sequences transmitted on different subcarriers, the weaker their mutual influence. Therefore, we believe that the attention mechanism of the TransDetector could well suppress the influence of ICI, which directly explains why the more input data of the network, the better the performance is in Section~\ref{Section:SimulationResults}-A, i.e., the network can focus on and reduce more interference sources at the same time.

\section{Conclusions} \label{Section:Conclusions}
In this paper, we have proposed a Transformer-based detector for UWA-DOFDM signal detection, referred to as the TransDetector, which can effectively mitigate ICI and noise in the received signal. In the TransDetector, three techniques, i.e., the T-PE, the DNS, and the I-MHA, have been innovatively designed to enhance the performance of the standard Transformer. Extensive experimental results on both the simulation channel and the realistic underwater channel have shown the superiority of the proposed TransDetector over the classical $\mathcal{X}$-FFT and the DNNDetector. For example, compared with the $\mathcal{X}$-FFT and the DNNDetector, the TransDetector reduces the BER by $47.44\%$ and $33.49\%$ in the case of $\text{SNR}=20~\text{dB}$, respectively, when the realistic channel is applied.


\bibliographystyle{IEEEtran}
\bibliography{myRef}

\end{document}